\documentclass[aps,pre,twocolumn,superscriptaddress,floatfix,tightenlines,showpacs,notitlepage]{revtex4-1}
\usepackage{graphicx}
\usepackage{bm}
\usepackage[utf8]{inputenc}
\usepackage[english]{babel}
\usepackage{amsfonts}
\usepackage{color}
\usepackage{amsmath,amssymb}    
\usepackage{epsfig}
\usepackage{subfigure}  
\usepackage{hyperref}   
\usepackage{amssymb}
\usepackage{soul}
\usepackage[section]{placeins}
\usepackage{url}
\usepackage{ulem}
\AtBeginDocument{%
    \newwrite\bibnotes
    \def\bibnotesext{Notes.bib}
    \immediate\openout\bibnotes=\jobname\bibnotesext
    \immediate\write\bibnotes{@CONTROL{REVTEX41Control}}
    \immediate\write\bibnotes{@CONTROL{%
    apsrev41Control,author="08",editor="1",pages="1",title="0",year="1"}}
     \if@filesw
     \immediate\write\@auxout{\string\citation{apsrev41Control}}%
    \fi
}%

\newcommand{\ictsaddress}{International Centre for Theoretical Sciences, Tata Institute of Fundamental Research, Bangalore 560089, India}
\newcommand{\ocaaddress}{Universit\'e C\^ote d’Azur, CNRS, LJAD, Nice 06100, France}
\newcommand{\iitbaddress}{Department of Chemical Engineering, Indian Institute of Technology Bombay, Mumbai 400076, India}
\newcommand{\Minnesotaaddress}{School of Physics and Astronomy, University of Minnesota, Minneapolis, Minnesota 55455, USA}

\begin{document}
\title{Elasto-inertial Chains in a Two-dimensional Turbulent Flow}

\author{Rahul Singh}
\email{rahul.singh@icts.res.in}
\affiliation{\ictsaddress}

\author{Mohit Gupta}
\email{mohit.gupta9607@gmail.com}
\affiliation{\ictsaddress}
\affiliation{\Minnesotaaddress}

\author{Jason R. Picardo}
\email{picardo21@gmail.com}
\affiliation{\ictsaddress}
\affiliation{\iitbaddress}

\author{Dario Vincenzi}
\email{dario.vincenzi@unice.fr}
\altaffiliation{Also: Associate, International Centre for Theoretical Sciences, Tata Institute of Fundamental Research, Bangalore 560089, India}
\affiliation{\ocaaddress}

\author{Samriddhi Sankar Ray}
\email{samriddhisankarray@gmail.com}
\affiliation{\ictsaddress}


\begin{abstract}
The interplay of inertia and elasticity is shown to have a significant impact on the transport of filamentary objects, modelled by bead-spring chains, in a two-dimensional turbulent flow. We show how elastic interactions amongst inertial 
beads result in a non-trivial sampling of the flow, ranging from entrapment within vortices to preferential sampling of straining regions. This behavior is quantified as a function 
of inertia and elasticity and is shown to be very different from free, non-interacting 
heavy particles, as well as inertialess chains 
[Picardo \textit{et al.}, Phys. Rev. Lett. {\bf 121}, 244501 (2018)]. In addition, by considering two limiting cases, of a heavy-headed and a uniformly-inertial chain, we illustrate the critical role played by the mass distribution of such extended objects in their turbulent transport.
\end{abstract}
\maketitle

The study of the dynamics of a long filamentary object in a turbulent
flow is fairly recent. 
In particular, studies on the deformation \citep{Brouzet_polymer,Verhille_3dconf_fiber,Gay2018,Marchioli} and buckling \citep{Bec_Chain} of 
fibers, as well as on
their usefulness as a probe for the statistical properties
of a turbulent flow~\citep{Brandt_fiber,Mazzino2019}, have lead to the development of new ideas in the area of
turbulent transport which go beyond the spherical point-particle approximation. 

The dynamics of such filamentary objects becomes particularly intriguing when their length extends beyond the dissipation scale of the flow; 
the mean flow velocity sampled by the object is then dependent on its instantaneous shape, 
and this couples translation to flow-induced deformation. Recently, the work of Picardo {\it et al.}~\citep{PicardoPRL}, which
modelled extensible elastic chains as strings of inertialess tracers linked by springs, revealed
a new mechanism by which such objects preferentially sample the flow: unlike a
collection of non-interacting tracers, which distribute homogeneously,
these chains
preferentially sample the vortical regions of a two-dimensional turbulent flow,
with a non-trivial dependence on the elasticity (quantified by the Weissenberg number $\rm Wi$) and typical inter-bead separations in the chain. 

This behavior of elastic chains is in contrast to the well-known preferential sampling of straining regions exhibited by non-interacting, heavy particles 
(whose finite inertia is
measured through the Stokes number ${\rm St}$). More popularly known as ``preferential concentration'', this phenomenon has been the subject of extensive research in the last decade~\citep{Bec_frac_clustering,Bec_collisions, chun_aerosol,Bec_Biferale_heavy_particle,Monchaux_pref_conc, Gustavsson_heavy_particles,Bec_gravity,good_inertial_settling}, motivated
by its relevance to a diverse range of physical processes, from transport in particle-laden sprays~\citep{ssahu2016} 
to collision-driven growth of droplets in warm clouds \citep{Grabowski2013,Bec2016}.

It is critical to appreciate that the mechanisms of preferential sampling by the elastic inertia-less chains and the 
non-interacting, inertial (heavy)
particles are fundamentally different. In the former, it
is the elasticity of the links which allows such chains to extend, coil up and be
trapped in vortices, whereas for the latter the dissipative
dynamics and centrifugal expulsion from vortices lead to particles concentrating in straining
regions. This contrast
naturally leads us to investigate the dynamics of heavy elastic chains, which serve as a model for extensible filamentary objects that possess both inertia and elasticity, as is the case in most physical situations.
In addition, by varying the masses of the particles that compose a chain, we can study the effect of an inhomogenous mass distribution
along the chain itself.

\begin{figure}
\includegraphics[width=1.0\columnwidth]{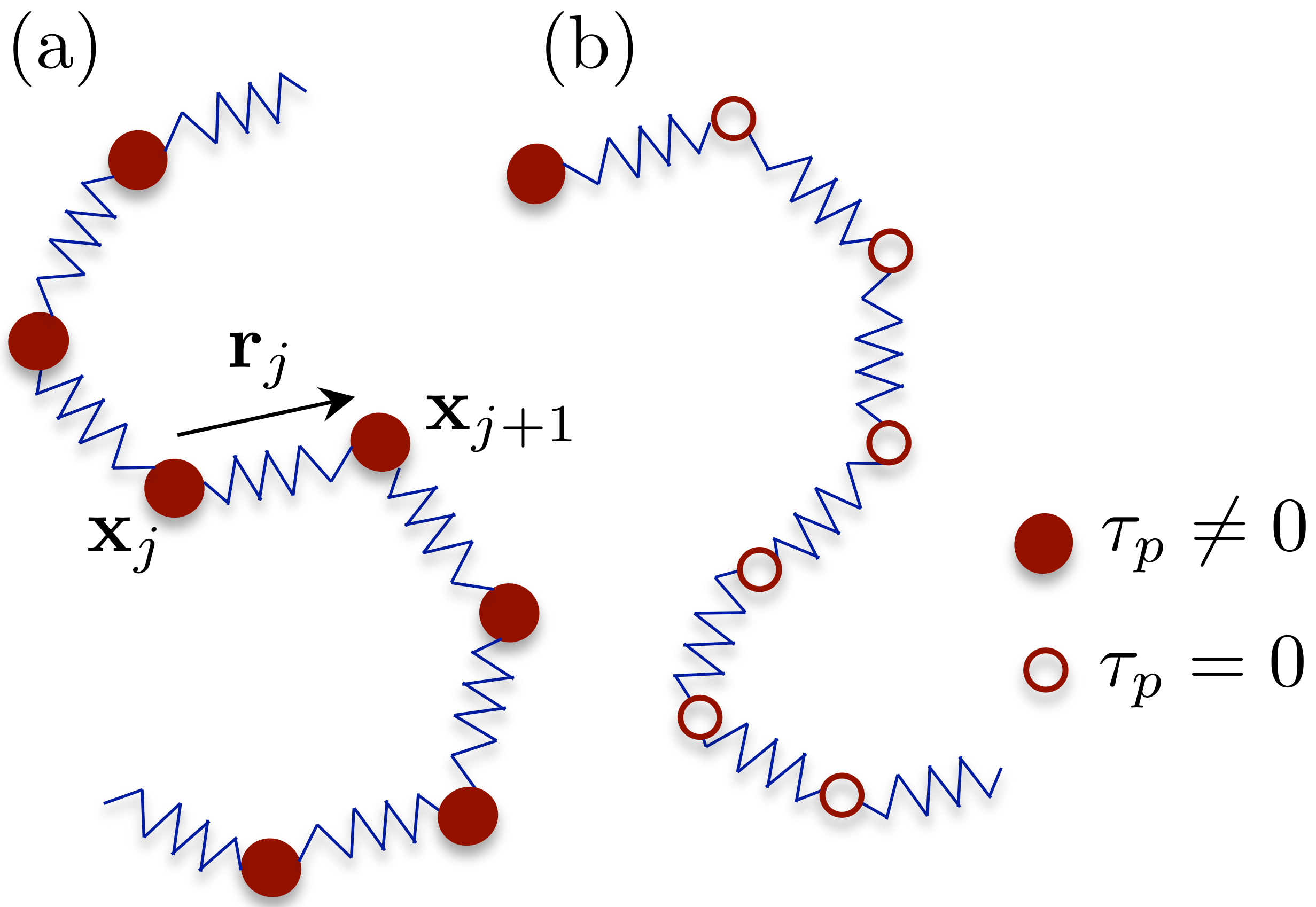}
\caption{A schematic of (a) a uniformly-inertial and (b) a heavy-headed chain illustrating the notation used in the text.}
\label{schematic}
\end{figure}

Towards this end, we consider
an \textit{elasto-inertial chain}, {\it i.e.} a sequence of heavy spherical particles---henceforth called beads---which are connected to their 
nearest neighbours through elastic (phantom) links [see Fig.~\ref{schematic}(a)] with an associated time scale 
$\tau_E$. The chain is composed of $N_b$ beads with positions ${\bf x}_j$, $1\leqslant j\leqslant N_b$,
and each bead, in turn, is characterised by the inertial relaxation time $\tau_p$ 
with which its velocity would relax to that of the fluid in the absence of elastic interactions with the neighboring beads.
By incorporating the drag stemming from the advecting fluid velocity field ${\bf u}$ and the elastic forces on each bead, 
it is easy to show that the equations of motion are most simply formulated in terms of the inter-bead separation vectors 
${\bf r}_j = {\bf x}_{j+1} - \textbf{x}_{j}$ with $1\leqslant j \leqslant N_b-1$ [see Fig.~\ref{schematic}(a)]:
\begin{widetext}
\begin{equation}
\tau_p\ddot{{\bf r}}_j = \left[{\bf u}({\bf x}_{j+1},t)-{\bf u}({\bf x}_j,t)- \dot{{\bf r}}_j\right]+ \sqrt{\frac{r^2_0}{2\tau_E}}\left [{\bm \xi}_{j+1}(t)-{\bm \xi}_j(t)\right ] 
+\frac{1}{4\tau_E}\left (f_{j-1}{\bf r}_{j-1}-2f_{j}{\bf r}_{j}+f_{j+1}{\bf r}_{j+1}\right ).
\label{eqr}
\end{equation}
\end{widetext}
Here, the ``link velocity'' is denoted as $\dot{{\bf r}}_j = \dot{{\bf x}}_{j+1}-\dot{{\bf x}}_{j}$  
and the  ``link acceleration'' as $\ddot{{\bf r}}_j$. We use the FENE (finitely extensible nonlinear elastic) interaction 
$f_j=(1-|\textbf{r}^2_j|/r^2_m)^{-1}$, where $r_m$ is the maximum inter-bead length, commonly used in polymer physics~\cite{Bird}. 
Finally, we budget for thermal 
fluctuations through independent white noises ${\bm \xi}_j(t)$ on each bead with an amplitude $r_0$ to set the equilibrium length of each link. Likewise, the equation of motion for the 
center of mass ${\bf x}_c$ is given by
\begin{equation}
\tau_p\ddot{{\bf x}}_c = \left(\frac{1}{N_b}\sum_{j=1}^{N_b}{{\bf u}({\bf x}_j,t)}-\dot{{\bf x}}_c\right)+\frac{1}{N_b}\sqrt{\frac{r^2_0}{2\tau_E}}\sum_{j=1}^{N_b}{\bm \xi}_j(t).
\label{eqcm}
\end{equation}
In Eqs.~\eqref{eqr} and ~\eqref{eqcm}, we have taken 
$\tau_p$, $\tau_E$, and $r_0$ to be identical for all beads and links. Thus we obtain a uniformly-inertial chain, which will be the main focus of our study. However, to explore the role of the mass-distribution of the chain, we also consider a second case, in which 
all the inertia is concentrated in a single heavy end-bead ($j = 1$), with the remainder of the chain composed of $N_b - 1$ inertia-less beads, as 
illustrated in Fig.~\ref{schematic}(b).
Such a ``heavy-headed chain" pits the inertia of the head-bead directly against the elasticity of its tail, and serves as an ideal candidate to illustrate the effects of these competing forces.
The equations of motion for the inter-bead links of such a heavy-headed chain are a specific instance of 
Eqs.~\eqref{eqr} and \eqref{eqcm}, and are given by:
\begin{widetext}
\begin{equation}
\dot{{\bf r}}_j = {\bf u}({\bf x}_{j+1},t)-{\bf u}({\bf x}_j,t)+\sqrt{\frac{r^2_0}{2\tau_E}}\left[{\bm \xi}_{j+1}(t)-{\bm \xi}_j(t)\right] 
-\frac{1}{4\tau_E}(2f_{j}{\bf r}_{j}-f_{j-1}{\bf r}_{j-1}-f_{j+1}{\bf r}_{j+1}) \quad \forall j \neq 1
\label{eqrhead}
\end{equation}
\end{widetext}
and for $j=1$:
\begin{equation}
\dot{{\bf r}}_1 = {\bf u}({\bf x}_2,t)+\sqrt{\frac{r^2_0}{2\tau_E}}{\bm \xi}_{2}(t) +\frac{1}{4\tau_E}(f_{2}{\bf r}_{2}-f_{1}{\bf r}_{1})-\dot{{\bf x}}_1.
\label{eqr1}
\end{equation}
The center of mass, which coincides with the head bead, obeys:
\begin{equation}
\tau_p\ddot{{\bf x}}_1=\left({\bf u}({\bf x}_1,t)-\dot{{\bf x}}_1\right)+\sqrt{\frac{r^2_0}{2\tau_E}}{\bm \xi}_1(t)+\frac{1}{4\tau_E}f_{1}{\bf r}_{1}.
\label{eqrcmh}
\end{equation}
Equations~\eqref{eqr} to~\eqref{eqrcmh} complete the description of the dynamics of our two types of elasto-inertial chains. However, it is important 
to keep in mind that although we have gone beyond the tracer approximation, this model does not incorporate other realistic effects such as 
hydrodynamic interactions amongst beads (which can become important when the inter-bead separation shrinks), and the beads themselves 
are assumed to be small enough---consistent with the linear Stokes drag model---not to affect the fluid flow field. Moreover, our elastic links do not
offer any resistance to bending, which ought to be important in any realistic modelling of fibers, for example.
The focus of this study is rather on the interplay between elasticity and inertia.

\begin{figure*}
\includegraphics[width=1.0\textwidth]{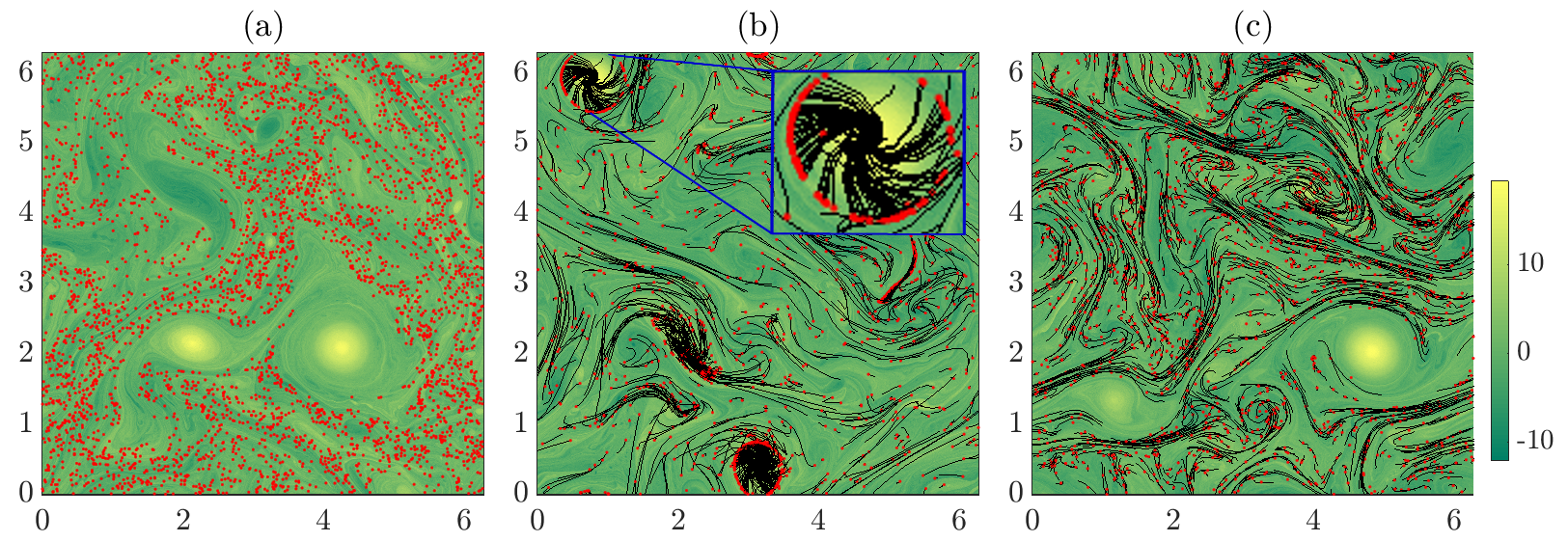}
\caption{Representative snapshots of a randomly chosen subset of (a) non-interacting inertial particles, as well as (b) heavy-headed chains and (c) uniformly-inertial 
chains overlaid on the vorticity field. The center-of-mass of the chains are shown by red dots [like the free particles in panel (a)] 
and the chain itself by black lines. The inset in panel (b) shows a zoomed-in view of the vortex located near the top-left corner of this panel. We show results for ${\rm St} = 0.14$ and 
${\rm Wi} = 1.38$ [for panels (b) and (c)].}
\label{com}
\end{figure*}

We immerse these chains in a two-dimensional, homogeneous, isotropic turbulent velocity field ${\bf u}$, which is obtained through direct numerical simulations (DNSs) of the incompressible ($\nabla\cdot {\bf u} = 0$) Navier-Stokes equation:
 \begin{equation}
\partial_t {\bf u} + ({\bf u} \cdot \nabla){\bf u} = -\nabla p + \nu \nabla^2 {\bf u} + {\bf f} - \mu {\bf u}.
\label{eq:NS}
\end{equation}
Two-dimensional flows are particularly useful for investigating the competing effects of elasticity and inertia 
because of long-lived, coherent vortical structures, as we shall see later. 
We use a standard pseudo-spectral method to solve Eq.~\eqref{eq:NS} on a 2$\pi$ square periodic domain with 
$N^2 = 1024^2$ collocation points. We drive the flow to a turbulent, statistically steady state with an external forcing 
${\bf f} = -F_0{\rm sin}(k_fx) {\bf e_y}$, where $F_0 $ is the forcing amplitude and $k_f$ sets the energy-injection and typical vortex 
scale $l_f = 2\pi k_f^{-1}$. The energy at large scales (due to an inverse cascade) is 
damped out by using an Ekmann term \citep{prasad2011,Boffetta-Ann-Rev} with the coefficient of friction  $\mu=10^{-2}$. The flow is characterized by the large eddy-turn-over time scale 
$\tau_f = l_f/\sqrt{2E} $ and the short time scale $\tau_\eta = 1/\sqrt{2\langle\omega^2\rangle}$
associated with enstrophy dissipation, 
where $E$ is the mean kinetic energy of the flow and $\langle\omega^2\rangle$ is the mean enstrophy. We set the coefficient of kinematic viscosity 
$\nu=1\times 10^{-6}$, $k_f=5$, and $F_0=0.2$, giving $\tau_f=1.45$ and $\tau_\eta=0.35$.

We seed the turbulent flow with $5\times10^4$ chains, whose dynamics, determined by Eqs.~\eqref{eqr} and~\eqref{eqcm}
or~\eqref{eqrhead} to~\eqref{eqrcmh}, are numerically 
integrated by a second-order Runge-Kutta scheme. Each chain consists of $N_b = 10$ beads 
and, hence, has a maximum length of 
$L_m=(N_b-1)r_m = 1.25$ (comparable to the forcing scale $l_f$) as well as an equilibrium link length $r_0 = 0.004$. 
The dynamics of the chain is controlled entirely by  
its elasticity and inertia, conveniently described by the 
Weissenberg number ${\rm Wi} = \tau_{\rm chain}/\tau_f$, where $\tau_{\rm chain}=6\tau_{E}/(N_b(N_b+1))$ provides an estimate of the effective relaxation time of the entire chain \citep{Collins2007}, 
and the Stokes number ${\rm St} = \tau_p/\tau_\eta$, respectively. 
We use several values of ${\rm Wi}$ and ${\rm St}$ to explore the different regimes 
in the behavior of our elasto-inertial chains. However, 
 we focus on
the dynamics of individual chains and, therefore, ignore inter-chain interactions;
the use of a large number of chains, given that they are non-interacting, ensures that we obtain reliable statistics on the measurements that 
we make. 

We begin our study by asking: (i) are elasto-inertial chains really
different from non-interacting inertial particles and (ii) does the use of inertial
beads, instead of tracer ones as in Ref.~\citep{PicardoPRL},  modify the dynamics qualitatively? 
We answer these questions first in the context of the heavy-headed chains [Eqs.~\eqref{eqrhead}--\eqref{eqrcmh}], 
where the competing influences of elasticity and inertia are most easily illustrated.

In Figs.~\ref{com}(a) and~\ref{com}(b), we show representative snapshots of (a) non-interacting
inertial particles and (b) heavy-headed chains (the inertial head bead is shown
in red, while the inertia-less filamentary tail is in black) in a
two-dimensional turbulent flow. The underlying vorticity fields, on which for clarity a random subset of 
particles or chains is overlaid, are different realisations of the same
statistically steady flow. We use ${\rm St} = 0.14$ for 
the non-interacting particles as well as for the
head bead of the chains, which have elastic tails with ${\rm Wi}=1.38$.
As expected, the
inertial non-interacting particles preferentially concentrate in the
straining zones of the flow. The behavior of the heavy-headed chains in Fig~\ref{com}(b),  though,
is in stark contrast to this, 
and also differs from the dynamics of the 
inertia-less elastic chains studied in Ref.~\citep{PicardoPRL}. Indeed, in the absence of inertia, elastic chains coil up into vortices and shrink down to tracer-like objects, which then continue to reside inside  vortices. 
The inertia-less chains in straining regions, however, are rapidly stretched out until they depart from the underlying straining flow and encounter a vortex, where straining is weak. 
Thus, inertia-less chains get preferentially trapped inside vortices (see \cite{Youtube_tr} for a movie depicting this behaviour), and 
snapshots such as those in Fig.~\ref{com} would show them to be located well 
inside the core of vortices~\citep{PicardoPRL}.

\begin{figure*}
\includegraphics[width=1\textwidth]{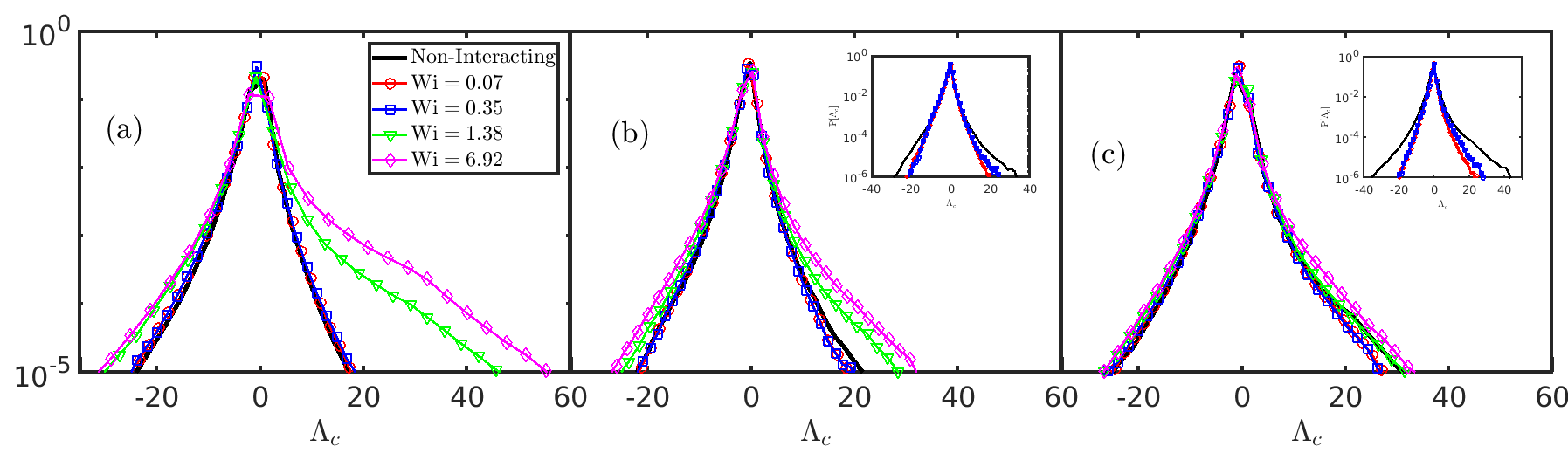}
\caption{Probability distribution functions of $\Lambda_c$ measured for uniformly-inertial chains with (a) ${\rm St} = 0.14$, (b) ${\rm St} = 0.85$ and (c) ${\rm St} = 2.84$. Curves are plotted for different degrees of elasticity of the chains, as well as for non-interacting 
inertial particles (see legend). The insets of panels (b) and (c) show the same distributions for the heavy-headed chains, but for only two values of ${\rm Wi}$.}
\label{pdf-lambda-inertial}
\end{figure*}

Returning to the snapshot of the heavy-headed chains in Fig.~\ref{com}(b), we see that a majority of them overlap with vortical regions, while remaining elongated, unlike inertia-less chains. 
The head beads, which are inertial, live on the periphery of the
vortices, while their filamentary, inertia-less elastic tails are pinned to the vortex cores, tracing out a Ferris-wheel pattern. This is especially clear when 
we look at the arrangement of the chains in and around the vortex visible in the top left 
corner of Fig.~\ref{com}(b) (the inset shows a zoomed-in view of this vortex). The filament, being
pinned to the core of the vortex  because of elasticity (through the mechanism identified in Ref.~\citep{PicardoPRL}), competes
with the centrifugal force on the head inertial bead which pushes them out of the vortex:
it is this competition between the two effects which manifests itself in the
head beads encircling the edge of the vortices. We refer the reader to 
 \cite{Youtube_hh} for a movie showing the motion of these chains 
in the flow, which illustrates this phenomenon---especially the Ferris-wheel pattern---clearly.

These results show that combining an inertial particle with an elastic tail gives rise to dynamics which are very different from that of either a free inertial particle or a purely elastic chain.
Such a competition between inertia and elasticity, which dictates the behaviour of heavy-headed chains, also impacts the dynamics of uniformly-inertial chains [Eqs.~\eqref{eqr} and \eqref{eqcm}], 
but in a less obvious manner. In Fig.~\ref{com}(c), which presents a snapshot of uniformly-inertial chains (see \cite{Youtube_ui} for a movie of the time-evolution of these chains), the core of the strongest vortices are evacuated, because of
centrifugal forces, while the weaker vortices are still occupied by partially coiled inertial chains.
The stark difference between panels (b) and (c) of Fig.~\ref{com}, provides a vivid illustration of the importance of the mass distribution of such long objects. For a better appreciation of these effects, we now turn to a more quantitative measurement of the sampling behavior of elasto-inertial chains.

A natural way to quantify the relative sampling of vortical and straining regions is to measure the
(Lagrangian) Okubo-Weiss parameter $\Lambda_c =
\frac{\omega_c^2-\sigma_c^2}{4\langle\omega^2\rangle}$ at the center of mass of
the chains along their trajectories. The vorticity $\omega_c$ and the strain
rate $\sigma_c$ are measured at the center of mass and normalized by the mean 
enstrophy $\langle\omega^2\rangle$ of the flow.  The
sign of this parameter is a signature of the local geometry of the flow: $\Lambda_c > 0$ implies that the center of mass lies in a vortical
region, while $\Lambda_c < 0$ is indicative of a straining zone.  As is obvious
from the definition of the Okubo-Weiss parameter, extremely small values
correspond to regions with comparable amounts of vorticity  and straining.

In Fig.~\ref{pdf-lambda-inertial}, we show the plots of the probability distribution function (PDF) of the
Okubo-Weiss parameter $\Lambda_c$ for non-interacting inertial particles and uniformly-inertial chains. We consider four values of ${\rm Wi}$ (0.07, 0.35, 1.38, 6.92) for each of three Stokes numbers: (a) ${\rm St} = 0.14$, (b) ${\rm St} = 0.85$, and (c) ${\rm St} = 2.84$.

For small, but still nonzero Stokes numbers [see
Fig.~\ref{pdf-lambda-inertial}(a)], non-interacting inertial particles have a
distribution of $\Lambda_c$ which is negatively skewed, 
indicating a preferential sampling of straining regions \cite{Dhruba2018}. 
(Note that uniformly distributed non-interacting \textit{tracers} would 
show a positively skewed PDF of $\Lambda_c$ 
owing to the presence of intense coherent vortices in the flow~\citep{prasad2011,Gupta2015}.) 
However, for a
chain of inertial beads with the same Stokes number, the effect of the elasticity draws the
chain (defined by its center of mass) towards more vortical regions.
This is clearly seen in the widening of the right tails of the PDF as ${\rm Wi}$ increases [Fig.~\ref{pdf-lambda-inertial}(a)].

\begin{figure*}
\includegraphics[width=\textwidth]{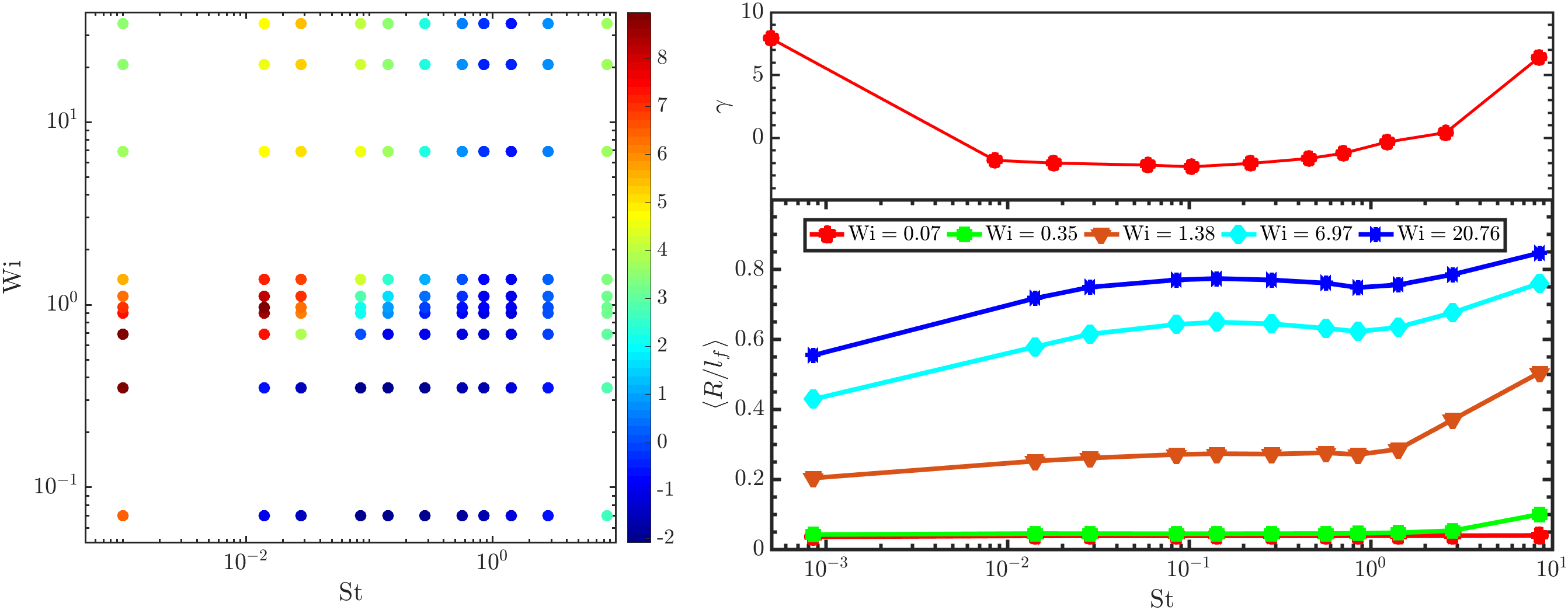}
\caption{(a) Pseudo-color plots of the skewness $\gamma$ of the distribution of $\Lambda_c$, in the ${\rm St}-{\rm Wi}$ plane, for uniformly-inertial chains. For comparison, the same skewness obtained for non-interacting particles is shown in panel (b). We also show, in panel (c), the average normalised 
length of the inertial chains as a function of ${\rm St}$ for a few representative values of ${\rm Wi}$.}
\label{skew-lambda-heavy}
\end{figure*}

This effect of elasticity persists, qualitatively, as ${\rm St}$ is increased to intermediate values [Fig.~\ref{pdf-lambda-inertial}(b)], but is considerably weaker. 
On the one hand, the increasing centrifugal forces acting on the chains counteract the tendency of elasticity to trap them into vortices, causing 
the large-${\rm Wi}$ PDFs to show less positively-skewed tails. On the other hand, the non-interacting inertial particles begin to decorrelate from the flow and start to distribute more uniformly which causes the corresponding PDFs of $\Lambda_c$
to become more positively skewed. The net result is that the effect of ${\rm Wi}$ weakens, 
and eventually for large ${\rm St}$ the PDFs of $\Lambda_c$ become nearly independent of elasticity [Fig.~\ref{pdf-lambda-inertial}(c)].

Do we expect the heavy-headed chains to exhibit a similar behavior? The answer is no. This is because the heavy head bead, which coincides with the centre of mass, is held at the periphery of vortices by their elastic tails [Fig.~\ref{com}(c)]. Thus, the centre-of-mass is not allowed to sample either intense vortical or straining regions. We, therefore, expect the corresponding PDFs of $\Lambda_c$
to have relatively narrow tails, for both positive and negative values of $\Lambda_c$. These ideas are confirmed by the insets of Fig.~\ref{pdf-lambda-inertial}, which show the PDFs of $\Lambda_c$ for heavy-headed chains, for ${\rm St} = 0.85$ [panel (b)] and 
${\rm St} = 2.84$ [panel (c)]. For clarity, we only present representative results, in each case, for ${\rm Wi} = 0.07$ and 0.35. 
We have checked that the PDFs of $\Lambda_c$ for other values of ${\rm Wi}$ are similar to the ones 
shown and their tails remain bounded by the corresponding distribution of the
non-interacting particles. As we conjectured, the widths of these PDFs are always 
smaller than that of non-interacting inertial particles as well as uniformly-inertial chains. 
Of course, in the limit of
${\rm Wi} \ll 1$ or ${\rm St} \gg 1$, elasto-inertial chains show a dynamics much closer to 
that exhibited by a free inertial particle. 

The PDFs shown in Fig.~\ref{pdf-lambda-inertial} do suggest a non-trivial 
behavior of the skewness in the distribution of $\Lambda_c$ when compared with the non-interacting inertial 
particles, or even with the inertia-less elastic chains in Ref.~\citep{PicardoPRL}. An obvious way
to quantify this complex dynamics is to measure the skewness of these distributions 
$\gamma = \langle (\Lambda_{c}-\mu)^3 \rangle/{{\langle (\Lambda_{c}-\mu)^2 \rangle}}^{3/2}$, where $\mu = \langle \Lambda_c \rangle$,
as a function of both the Stokes and Weissenberg numbers. 
For comparison, let us first consider the plot of $\gamma$ vs ${\rm St}$ for non-interacting inertial particles 
shown in Fig.~\ref{skew-lambda-heavy}(b). Here, $\gamma$ is negative over intermediate values of ${\rm St}$, where there is evidence of preferential 
concentration (e.g., Ref.~\citep{Bec_Biferale_heavy_particle}), 
whereas for ${\rm St} \to 0$ or ${\rm St} \gg 1$, the homogeneous 
distribution of these particles ensures that  $\gamma$ is positive, as for tracers. When these same inertial particles are 
coupled elastically in the form of a chain, the nature of preferential sampling changes, resulting in a non-trivial 
dependence of $\gamma$ on ${\rm St}$ and ${\rm Wi}$. 


In Fig.~\ref{skew-lambda-heavy}(a), we show a pseudo-color plot of the skewness $\gamma$ in the 
${\rm St}-{\rm Wi}$ plane for uniformly-inertial chains.
We see that chains which stretch 
marginally, i.e., for ${\rm Wi} \ll 1$, show qualitatively the same dependence of $\gamma$ on ${\rm St}$ as that of non-interacting particles [Fig.~\ref{skew-lambda-heavy}(b)]. 
However, when  ${\rm Wi} \gtrsim 1$, the chains are stretched out 
and, hence, start coiling inside vortices which leads
to positive values of $\gamma$ for small ${\rm St}$ and a significant shrinking of the range of ${\rm St}$  for which $\gamma < 0$.

Is the difference between an inertia-less and a uniformly-inertial chain as strong as
that between non-interacting and elastically-interacting inertial particles discussed
above? This question is answered when we compare the behavior of $\gamma$ vs
${\rm Wi}$ for ${\rm St}$ around unity with the case of ${\rm St} \to 0$ in
Fig.~\ref{skew-lambda-heavy}(a). Indeed, for nonzero Stokes numbers, $\gamma$
has a sensitive dependence on the degree of stretching and goes from negative
to positive values with increasing ${\rm Wi}$. In contrast, when ${\rm St}\to 0$, $\gamma$
remains strictly positive while showing a non-monotonic dependence on ${\rm Wi}$, as
has been shown in Ref.~\citep{PicardoPRL} and is further discussed below.  In
the opposite limit of very large ${\rm St}$, $\gamma$ is nearly independent of ${\rm Wi}$,
in accordance with the PDFs of $\Lambda_c$ shown in
Fig.~\ref{pdf-lambda-inertial}(c).  

Our arguments, so far, are largely based on the stretching of the chains. It
is, therefore, essential to check if stretching indeed happens in the way we
suggest. In Fig.~\ref{skew-lambda-heavy}(c) we show representative plots of the
average length of the chains $\langle R \rangle$, normalised by $l_f$, as a
function of the Stokes number, for different values of ${\rm Wi}$.  For chains with a
negligible ${\rm Wi}$, there is hardly any evidence of stretching. However, as soon
as ${\rm Wi}$ is nonzero, the chains start stretching much
more, in a manner that depends non-trivially on the inertia of the beads. 

The variation of the chain length with ${\rm St}$ occurs in two distinct regimes, which are most clearly visible for the large-${\rm Wi}$ case in Fig.~\ref{skew-lambda-heavy}(c). For ${\rm St}\lesssim1$, inertial chains preferentially sample straining regions of the flow and, therefore, are stretched out more than a non-inertial chain, which has a much higher probability of being coiled into vortices, where it can relax to small lengths~\citep{PicardoPRL}. As ${\rm St}$ 
approaches unity, however, the heavy chains begin to decorrelate from the flow and the degree of sampling of straining regions reduces [the same is seen for non-interacting particles in Fig.~\ref{skew-lambda-heavy}(b)]. This is why the chain lengths, especially for large ${\rm Wi}$, 
show a weak local maximum 
for intermediate values of ${\rm St}\approx 0.1$, 
where the preferential sampling is strong [see Fig.~\ref{skew-lambda-heavy}(b)].
In the second regime, of ${\rm St} > 1$, the chain lengths again show an increasing trend. This is because, unlike small-${\rm St}$ beads, 
which are well correlated with the flow, the velocity differences between large-${\rm St}$ beads do not scale with their separation---even beads that are close to each other can have very different velocities. Therefore, it is more difficult for the elastic links to keep such heavy inertial beads together and the chains elongate as ${\rm St}$ increases.

\begin{figure}
\includegraphics[width=\columnwidth]{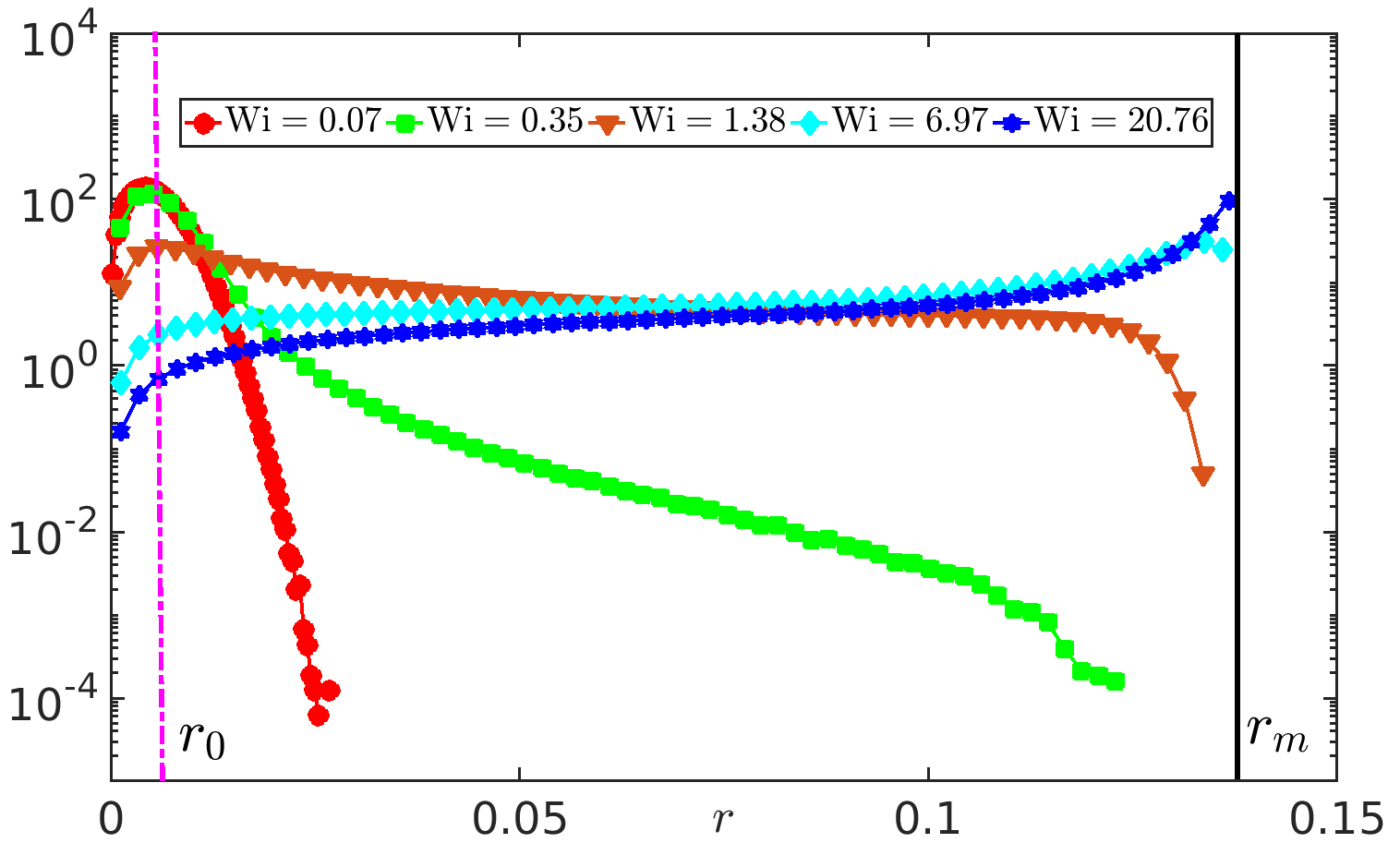}
\caption{Probability distribution function of the inter-bead separation $r$ for uniformly-inertial (${\rm St} = 0.14$) chains, for different values of ${\rm Wi}$ (see legend).
The dashed-dot (red) and the solid (black) vertical lines correspond to the equilibrium $r_0$ and maximum $r_m$ lengths of the links. With increasing elasticity, we 
see the distribution develop broader tails and eventually, at large values of ${\rm Wi}$, peaks near $r_m$.}
\label{link}
\end{figure}

The stretching of the chain is, of course, rooted in the distribution of the lengths of individual links. In Fig.~\ref{link}, we show a 
representative plot of this PDF for ${\rm St} = 0.14$ and for different values of ${\rm Wi}$. We notice that when ${\rm Wi}$ is very small, the PDF 
is narrow with a peak, as expected, near to the equilibrium link-length $r_0$. With increasing ${\rm Wi}$, the shape of this distribution undergoes a dramatic change, 
reflecting the stretching caused by the flow; an effect which is enhanced by the fact that the inertial beads preferentially sample regions of straining. The PDF initially
develops broader tails, but which, for ${\rm Wi}\lesssim 1$, are still far from the cut-off $r_m$ imposed in our model. As ${\rm Wi}$ increases beyond unity, however, 
the distribution starts getting flatter and eventually peaks near to
$r = r_m$. For such large-${\rm Wi}$ chains, the typical inter-bead separation is of the order of the vortex size; this limits the ability of the chains to coil into vortices, as successive beads can no longer simultaneously encounter the same vortex (finite $N_b$ acts a restriction on deformability). As explained previously for inertia-less chains~\citep{PicardoPRL}, this is the reason why $\gamma$, although strictly positive for ${\rm St} \to 0$, shows a peak for 
${\rm Wi} \approx 1$
and gradually decreases at large values of ${\rm Wi}$ [Fig.~\ref{skew-lambda-heavy}(a)]. 
For ${\rm St} \gtrsim 0.1$, 
however, the centrifugal forces acting on the beads prevent excessive entrapment of chains into vortices, thus eliminating the local maximum in $\gamma$ and resulting in a more gradual and monotonic increase with ${\rm Wi}$.

Inertia and elasticity are two fundamental properties of extended objects.
This work has shown, in the context of a model elasto-inertial chain, how the
interplay of these two features results in a non-trivial sampling of a
turbulent flow, thereby adding to recent studies on the turbulent transport of
objects with internal degrees of freedom. In particular, our results highlight
the significant differences between inertia-less and elasto-inertial chains, on
one hand, and free and elastically interacting inertial particles, on the
other. Moreover, the two limiting cases of a heavy-headed chain and a uniformly
heavy one, studied here, reveal the significant impact that the distribution of
mass can have on the turbulent transport of extended objects.

The elasto-inertial chain provides a simple way to account for the simultaneous effects of inertia, elasticity and fluid drag in models of filamentary objects, 
such as algae in marine environments, bio-filaments 
(actin and microtubules, for instance), and swimming microorganisms. In connection with the latter, recent studies have shown how the inertia of small active agents can aid in their clustering \citep{Breier2018} and flocking \citep{SSRay2018,Choudhary2015} in turbulent flows. The possible role of elasticity, however, remains to be addressed. Of course, such applications would require the consideration of additional effects, such as an active swimming velocity and inter-chain interactions. 
Nevertheless, the basic ideas elucidated here, in particular the competing effects of inertia and elasticity on preferential sampling, should remain relevant and help pave the way for future studies.

\begin{acknowledgments} 
SSR acknowledges DST (India) project ECR/2015/000361 for financial support. 
JRP, SSR, and DV  acknowledge the support of the Indo-French Centre for Applied Mathematics. The simulations were performed on the ICTS clusters {\it Mowgli}, {\it Tetris} and {\it Mario} as well as the work stations from the project ECR/2015/000361: {\it Goopy} and {\it Bagha}.  
\end{acknowledgments} 

\bibliography{ref_turb_chains}

\begin{thebibliography}{31}%
\makeatletter
\providecommand \@ifxundefined [1]{%
 \@ifx{#1\undefined}
}%
\providecommand \@ifnum [1]{%
 \ifnum #1\expandafter \@firstoftwo
 \else \expandafter \@secondoftwo
 \fi
}%
\providecommand \@ifx [1]{%
 \ifx #1\expandafter \@firstoftwo
 \else \expandafter \@secondoftwo
 \fi
}%
\providecommand \natexlab [1]{#1}%
\providecommand \enquote  [1]{``#1''}%
\providecommand \bibnamefont  [1]{#1}%
\providecommand \bibfnamefont [1]{#1}%
\providecommand \citenamefont [1]{#1}%
\providecommand \href@noop [0]{\@secondoftwo}%
\providecommand \href [0]{\begingroup \@sanitize@url \@href}%
\providecommand \@href[1]{\@@startlink{#1}\@@href}%
\providecommand \@@href[1]{\endgroup#1\@@endlink}%
\providecommand \@sanitize@url [0]{\catcode `\\12\catcode `\$12\catcode
  `\&12\catcode `\#12\catcode `\^12\catcode `\_12\catcode `\%12\relax}%
\providecommand \@@startlink[1]{}%
\providecommand \@@endlink[0]{}%
\providecommand \url  [0]{\begingroup\@sanitize@url \@url }%
\providecommand \@url [1]{\endgroup\@href {#1}{\urlprefix }}%
\providecommand \urlprefix  [0]{URL }%
\providecommand \Eprint [0]{\href }%
\providecommand \doibase [0]{http://dx.doi.org/}%
\providecommand \selectlanguage [0]{\@gobble}%
\providecommand \bibinfo  [0]{\@secondoftwo}%
\providecommand \bibfield  [0]{\@secondoftwo}%
\providecommand \translation [1]{[#1]}%
\providecommand \BibitemOpen [0]{}%
\providecommand \bibitemStop [0]{}%
\providecommand \bibitemNoStop [0]{.\EOS\space}%
\providecommand \EOS [0]{\spacefactor3000\relax}%
\providecommand \BibitemShut  [1]{\csname bibitem#1\endcsname}%
\let\auto@bib@innerbib\@empty
\bibitem [{\citenamefont {Brouzet}\ \emph {et~al.}(2014)\citenamefont
  {Brouzet}, \citenamefont {Verhille},\ and\ \citenamefont
  {Le~Gal}}]{Brouzet_polymer}%
  \BibitemOpen
  \bibfield  {author} {\bibinfo {author} {\bibfnamefont {C.}~\bibnamefont
  {Brouzet}}, \bibinfo {author} {\bibfnamefont {G.}~\bibnamefont {Verhille}}, \
  and\ \bibinfo {author} {\bibfnamefont {P.}~\bibnamefont {Le~Gal}},\
  }\bibfield  {title} {\enquote {\bibinfo {title} {Flexible fiber in a
  turbulent flow: A macroscopic polymer},}\ }\href {\doibase
  10.1103/PhysRevLett.112.074501} {\bibfield  {journal} {\bibinfo  {journal}
  {Phys. Rev. Lett.}\ }\textbf {\bibinfo {volume} {112}},\ \bibinfo {pages}
  {074501} (\bibinfo {year} {2014})}\BibitemShut {NoStop}%
\bibitem [{\citenamefont {Verhille}\ and\ \citenamefont
  {Bartoli}(2016)}]{Verhille_3dconf_fiber}%
  \BibitemOpen
  \bibfield  {author} {\bibinfo {author} {\bibfnamefont {G.}~\bibnamefont
  {Verhille}}\ and\ \bibinfo {author} {\bibfnamefont {A.}~\bibnamefont
  {Bartoli}},\ }\bibfield  {title} {\enquote {\bibinfo {title} {3d conformation
  of a flexible fiber in a turbulent flow},}\ }\href {\doibase
  10.1007/s00348-016-2201-1} {\bibfield  {journal} {\bibinfo  {journal} {Exp.
  Fluids}\ }\textbf {\bibinfo {volume} {57}},\ \bibinfo {pages} {117} (\bibinfo
  {year} {2016})}\BibitemShut {NoStop}%
\bibitem [{\citenamefont {Gay}\ \emph {et~al.}(2018)\citenamefont {Gay},
  \citenamefont {Favier},\ and\ \citenamefont {Verhille}}]{Gay2018}%
  \BibitemOpen
  \bibfield  {author} {\bibinfo {author} {\bibfnamefont {A.}~\bibnamefont
  {Gay}}, \bibinfo {author} {\bibfnamefont {B.}~\bibnamefont {Favier}}, \ and\
  \bibinfo {author} {\bibfnamefont {G.}~\bibnamefont {Verhille}},\ }\bibfield
  {title} {\enquote {\bibinfo {title} {Characterisation of flexible fibre
  deformations in turbulence},}\ }\href@noop {} {\bibfield  {journal} {\bibinfo
   {journal} {Europhys. Lett.}\ }\textbf {\bibinfo {volume} {123}},\ \bibinfo
  {pages} {24001} (\bibinfo {year} {2018})}\BibitemShut {NoStop}%
\bibitem [{\citenamefont {Dotto}\ and\ \citenamefont
  {Marchioli}(2019)}]{Marchioli}%
  \BibitemOpen
  \bibfield  {author} {\bibinfo {author} {\bibfnamefont {D.}~\bibnamefont
  {Dotto}}\ and\ \bibinfo {author} {\bibfnamefont {C.}~\bibnamefont
  {Marchioli}},\ }\bibfield  {title} {\enquote {\bibinfo {title} {Orientation,
  distribution, and deformation of inertial flexible fibers in turbulent
  channel flow},}\ }\href@noop {} {\bibfield  {journal} {\bibinfo  {journal}
  {Acta Mech.}\ }\textbf {\bibinfo {volume} {230}},\ \bibinfo {pages} {597}
  (\bibinfo {year} {2019})}\BibitemShut {NoStop}%
\bibitem [{\citenamefont {Allende}\ \emph {et~al.}(2018)\citenamefont
  {Allende}, \citenamefont {Henry},\ and\ \citenamefont {Bec}}]{Bec_Chain}%
  \BibitemOpen
  \bibfield  {author} {\bibinfo {author} {\bibfnamefont {S.}~\bibnamefont
  {Allende}}, \bibinfo {author} {\bibfnamefont {C.}~\bibnamefont {Henry}}, \
  and\ \bibinfo {author} {\bibfnamefont {J.}~\bibnamefont {Bec}},\ }\bibfield
  {title} {\enquote {\bibinfo {title} {Stretching and buckling of small elastic
  fibers in turbulence},}\ }\href {\doibase 10.1103/PhysRevLett.121.154501}
  {\bibfield  {journal} {\bibinfo  {journal} {Phys. Rev. Lett.}\ }\textbf
  {\bibinfo {volume} {121}},\ \bibinfo {pages} {154501} (\bibinfo {year}
  {2018})}\BibitemShut {NoStop}%
\bibitem [{\citenamefont {Rosti}\ \emph {et~al.}(2018)\citenamefont {Rosti},
  \citenamefont {Banaei}, \citenamefont {Brandt},\ and\ \citenamefont
  {Mazzino}}]{Brandt_fiber}%
  \BibitemOpen
  \bibfield  {author} {\bibinfo {author} {\bibfnamefont {M.~E.}\ \bibnamefont
  {Rosti}}, \bibinfo {author} {\bibfnamefont {A.~A.}\ \bibnamefont {Banaei}},
  \bibinfo {author} {\bibfnamefont {L.}~\bibnamefont {Brandt}}, \ and\ \bibinfo
  {author} {\bibfnamefont {A.}~\bibnamefont {Mazzino}},\ }\bibfield  {title}
  {\enquote {\bibinfo {title} {Flexible fiber reveals the two-point statistical
  properties of turbulence},}\ }\href {\doibase 10.1103/PhysRevLett.121.044501}
  {\bibfield  {journal} {\bibinfo  {journal} {Phys. Rev. Lett.}\ }\textbf
  {\bibinfo {volume} {121}},\ \bibinfo {pages} {044501} (\bibinfo {year}
  {2018})}\BibitemShut {NoStop}%
\bibitem [{\citenamefont {Rosti}\ \emph {et~al.}(2019)\citenamefont {Rosti},
  \citenamefont {Olivieri}, \citenamefont {Banaei}, \citenamefont {Brandt},\
  and\ \citenamefont {Mazzino}}]{Mazzino2019}%
  \BibitemOpen
  \bibfield  {author} {\bibinfo {author} {\bibfnamefont {M.~E.}\ \bibnamefont
  {Rosti}}, \bibinfo {author} {\bibfnamefont {S.}~\bibnamefont {Olivieri}},
  \bibinfo {author} {\bibfnamefont {A.~A.}\ \bibnamefont {Banaei}}, \bibinfo
  {author} {\bibfnamefont {L.}~\bibnamefont {Brandt}}, \ and\ \bibinfo {author}
  {\bibfnamefont {A.}~\bibnamefont {Mazzino}},\ }\bibfield  {title} {\enquote
  {\bibinfo {title} {Flowing fibers as a proxy of turbulence statistics},}\
  }\href@noop {} {\bibfield  {journal} {\bibinfo  {journal} {Meccanica}\ ,\
  \bibinfo {pages} {in press}} (\bibinfo {year} {2019})}\BibitemShut {NoStop}%
\bibitem [{\citenamefont {Picardo}\ \emph {et~al.}(2018)\citenamefont
  {Picardo}, \citenamefont {Vincenzi}, \citenamefont {Pal},\ and\ \citenamefont
  {Ray}}]{PicardoPRL}%
  \BibitemOpen
  \bibfield  {author} {\bibinfo {author} {\bibfnamefont {J.~R.}\ \bibnamefont
  {Picardo}}, \bibinfo {author} {\bibfnamefont {D.}~\bibnamefont {Vincenzi}},
  \bibinfo {author} {\bibfnamefont {N.}~\bibnamefont {Pal}}, \ and\ \bibinfo
  {author} {\bibfnamefont {S.~S.}\ \bibnamefont {Ray}},\ }\bibfield  {title}
  {\enquote {\bibinfo {title} {Preferential sampling of elastic chains in
  turbulent flows},}\ }\href {\doibase 10.1103/PhysRevLett.121.244501}
  {\bibfield  {journal} {\bibinfo  {journal} {Phys. Rev. Lett.}\ }\textbf
  {\bibinfo {volume} {121}},\ \bibinfo {pages} {244501} (\bibinfo {year}
  {2018})}\BibitemShut {NoStop}%
\bibitem [{\citenamefont {Bec}(2003)}]{Bec_frac_clustering}%
  \BibitemOpen
  \bibfield  {author} {\bibinfo {author} {\bibfnamefont {J.}~\bibnamefont
  {Bec}},\ }\bibfield  {title} {\enquote {\bibinfo {title} {Fractal clustering
  of inertial particles in random flows},}\ }\href {\doibase 10.1063/1.1612500}
  {\bibfield  {journal} {\bibinfo  {journal} {Phys. Fluids}\ }\textbf {\bibinfo
  {volume} {15}},\ \bibinfo {pages} {L81--L84} (\bibinfo {year}
  {2003})}\BibitemShut {NoStop}%
\bibitem [{\citenamefont {Bec}\ \emph {et~al.}(2005)\citenamefont {Bec},
  \citenamefont {Celani}, \citenamefont {Cencini},\ and\ \citenamefont
  {Musacchio}}]{Bec_collisions}%
  \BibitemOpen
  \bibfield  {author} {\bibinfo {author} {\bibfnamefont {J.}~\bibnamefont
  {Bec}}, \bibinfo {author} {\bibfnamefont {A.}~\bibnamefont {Celani}},
  \bibinfo {author} {\bibfnamefont {M.}~\bibnamefont {Cencini}}, \ and\
  \bibinfo {author} {\bibfnamefont {S.}~\bibnamefont {Musacchio}},\ }\bibfield
  {title} {\enquote {\bibinfo {title} {Clustering and collisions of heavy
  particles in random smooth flows},}\ }\href {\doibase 10.1063/1.1940367}
  {\bibfield  {journal} {\bibinfo  {journal} {Phys. Fluids}\ }\textbf {\bibinfo
  {volume} {17}},\ \bibinfo {pages} {073301} (\bibinfo {year}
  {2005})}\BibitemShut {NoStop}%
\bibitem [{\citenamefont {Chun}\ \emph {et~al.}(2005)\citenamefont {Chun},
  \citenamefont {Koch}, \citenamefont {Rani}, \citenamefont {Ahluwalia},\ and\
  \citenamefont {Collins}}]{chun_aerosol}%
  \BibitemOpen
  \bibfield  {author} {\bibinfo {author} {\bibfnamefont {J.}~\bibnamefont
  {Chun}}, \bibinfo {author} {\bibfnamefont {D.~L.}\ \bibnamefont {Koch}},
  \bibinfo {author} {\bibfnamefont {S.~L.}\ \bibnamefont {Rani}}, \bibinfo
  {author} {\bibfnamefont {A.}~\bibnamefont {Ahluwalia}}, \ and\ \bibinfo
  {author} {\bibfnamefont {L.~R.}\ \bibnamefont {Collins}},\ }\bibfield
  {title} {\enquote {\bibinfo {title} {Clustering of aerosol particles in
  isotropic turbulence},}\ }\href {\doibase 10.1017/S0022112005004568}
  {\bibfield  {journal} {\bibinfo  {journal} {J. Fluid Mech.}\ }\textbf
  {\bibinfo {volume} {536}},\ \bibinfo {pages} {219–251} (\bibinfo {year}
  {2005})}\BibitemShut {NoStop}%
\bibitem [{\citenamefont {Bec}\ \emph {et~al.}(2007)\citenamefont {Bec},
  \citenamefont {Biferale}, \citenamefont {Cencini}, \citenamefont {Lanotte},
  \citenamefont {Musacchio},\ and\ \citenamefont
  {Toschi}}]{Bec_Biferale_heavy_particle}%
  \BibitemOpen
  \bibfield  {author} {\bibinfo {author} {\bibfnamefont {J.}~\bibnamefont
  {Bec}}, \bibinfo {author} {\bibfnamefont {L.}~\bibnamefont {Biferale}},
  \bibinfo {author} {\bibfnamefont {M.}~\bibnamefont {Cencini}}, \bibinfo
  {author} {\bibfnamefont {A.}~\bibnamefont {Lanotte}}, \bibinfo {author}
  {\bibfnamefont {S.}~\bibnamefont {Musacchio}}, \ and\ \bibinfo {author}
  {\bibfnamefont {F.}~\bibnamefont {Toschi}},\ }\bibfield  {title} {\enquote
  {\bibinfo {title} {Heavy particle concentration in turbulence at dissipative
  and inertial scales},}\ }\href {\doibase 10.1103/PhysRevLett.98.084502}
  {\bibfield  {journal} {\bibinfo  {journal} {Phys. Rev. Lett.}\ }\textbf
  {\bibinfo {volume} {98}},\ \bibinfo {pages} {084502} (\bibinfo {year}
  {2007})}\BibitemShut {NoStop}%
\bibitem [{\citenamefont {Monchaux}\ \emph {et~al.}(2012)\citenamefont
  {Monchaux}, \citenamefont {Bourgoin},\ and\ \citenamefont
  {Cartellier}}]{Monchaux_pref_conc}%
  \BibitemOpen
  \bibfield  {author} {\bibinfo {author} {\bibfnamefont {R.}~\bibnamefont
  {Monchaux}}, \bibinfo {author} {\bibfnamefont {M.}~\bibnamefont {Bourgoin}},
  \ and\ \bibinfo {author} {\bibfnamefont {A.}~\bibnamefont {Cartellier}},\
  }\bibfield  {title} {\enquote {\bibinfo {title} {Analyzing preferential
  concentration and clustering of inertial particles in turbulence},}\ }\href
  {\doibase https://doi.org/10.1016/j.ijmultiphaseflow.2011.12.001} {\bibfield
  {journal} {\bibinfo  {journal} {Int. J. Multiph. Flow}\ }\textbf {\bibinfo
  {volume} {40}},\ \bibinfo {pages} {1 -- 18} (\bibinfo {year}
  {2012})}\BibitemShut {NoStop}%
\bibitem [{\citenamefont {Gustavsson}\ and\ \citenamefont
  {Mehlig}(2016)}]{Gustavsson_heavy_particles}%
  \BibitemOpen
  \bibfield  {author} {\bibinfo {author} {\bibfnamefont {K.}~\bibnamefont
  {Gustavsson}}\ and\ \bibinfo {author} {\bibfnamefont {B.}~\bibnamefont
  {Mehlig}},\ }\bibfield  {title} {\enquote {\bibinfo {title} {Statistical
  models for spatial patterns of heavy particles in turbulence},}\ }\href
  {\doibase 10.1080/00018732.2016.1164490} {\bibfield  {journal} {\bibinfo
  {journal} {Adv. in Phys.}\ }\textbf {\bibinfo {volume} {65}},\ \bibinfo
  {pages} {1--57} (\bibinfo {year} {2016})}\BibitemShut {NoStop}%
\bibitem [{\citenamefont {Bec}\ \emph {et~al.}(2014)\citenamefont {Bec},
  \citenamefont {Homann},\ and\ \citenamefont {Ray}}]{Bec_gravity}%
  \BibitemOpen
  \bibfield  {author} {\bibinfo {author} {\bibfnamefont {J.}~\bibnamefont
  {Bec}}, \bibinfo {author} {\bibfnamefont {H.}~\bibnamefont {Homann}}, \ and\
  \bibinfo {author} {\bibfnamefont {S.~S.}\ \bibnamefont {Ray}},\ }\bibfield
  {title} {\enquote {\bibinfo {title} {Gravity-driven enhancement of heavy
  particle clustering in turbulent flow},}\ }\href {\doibase
  10.1103/PhysRevLett.112.184501} {\bibfield  {journal} {\bibinfo  {journal}
  {Phys. Rev. Lett.}\ }\textbf {\bibinfo {volume} {112}},\ \bibinfo {pages}
  {184501} (\bibinfo {year} {2014})}\BibitemShut {NoStop}%
\bibitem [{\citenamefont {Good}\ \emph {et~al.}(2014)\citenamefont {Good},
  \citenamefont {Ireland}, \citenamefont {Bewley}, \citenamefont {Bodenschatz},
  \citenamefont {Collins},\ and\ \citenamefont
  {Warhaft}}]{good_inertial_settling}%
  \BibitemOpen
  \bibfield  {author} {\bibinfo {author} {\bibfnamefont {G.~H.}\ \bibnamefont
  {Good}}, \bibinfo {author} {\bibfnamefont {P.~J.}\ \bibnamefont {Ireland}},
  \bibinfo {author} {\bibfnamefont {G.~P.}\ \bibnamefont {Bewley}}, \bibinfo
  {author} {\bibfnamefont {E.}~\bibnamefont {Bodenschatz}}, \bibinfo {author}
  {\bibfnamefont {L.~R.}\ \bibnamefont {Collins}}, \ and\ \bibinfo {author}
  {\bibfnamefont {Z.}~\bibnamefont {Warhaft}},\ }\bibfield  {title} {\enquote
  {\bibinfo {title} {Settling regimes of inertial particles in isotropic
  turbulence},}\ }\href {\doibase 10.1017/jfm.2014.602} {\bibfield  {journal}
  {\bibinfo  {journal} {J. Fluid Mech.}\ }\textbf {\bibinfo {volume} {759}},\
  \bibinfo {pages} {R3} (\bibinfo {year} {2014})}\BibitemShut {NoStop}%
\bibitem [{\citenamefont {Sahu}\ \emph {et~al.}(2016)\citenamefont {Sahu},
  \citenamefont {Hardalupas},\ and\ \citenamefont {Taylor}}]{ssahu2016}%
  \BibitemOpen
  \bibfield  {author} {\bibinfo {author} {\bibfnamefont {S.}~\bibnamefont
  {Sahu}}, \bibinfo {author} {\bibfnamefont {Y.}~\bibnamefont {Hardalupas}}, \
  and\ \bibinfo {author} {\bibfnamefont {A.~M. K.~P.}\ \bibnamefont {Taylor}},\
  }\bibfield  {title} {\enquote {\bibinfo {title} {Droplet–turbulence
  interaction in a confined polydispersed spray: effect of turbulence on
  droplet dispersion},}\ }\href {\doibase 10.1017/jfm.2016.169} {\bibfield
  {journal} {\bibinfo  {journal} {J. Fluid Mech.}\ }\textbf {\bibinfo {volume}
  {794}},\ \bibinfo {pages} {267–309} (\bibinfo {year} {2016})}\BibitemShut
  {NoStop}%
\bibitem [{\citenamefont {Grabowski}\ and\ \citenamefont
  {Wang}(2013)}]{Grabowski2013}%
  \BibitemOpen
  \bibfield  {author} {\bibinfo {author} {\bibfnamefont {W.~W.}\ \bibnamefont
  {Grabowski}}\ and\ \bibinfo {author} {\bibfnamefont {L.-P.}\ \bibnamefont
  {Wang}},\ }\bibfield  {title} {\enquote {\bibinfo {title} {Growth of cloud
  droplets in a turbulent environment},}\ }\href@noop {} {\bibfield  {journal}
  {\bibinfo  {journal} {Annu. Rev. Fluid Mech.}\ }\textbf {\bibinfo {volume}
  {45}},\ \bibinfo {pages} {293--324} (\bibinfo {year} {2013})}\BibitemShut
  {NoStop}%
\bibitem [{\citenamefont {Bec}\ \emph {et~al.}(2016)\citenamefont {Bec},
  \citenamefont {Ray}, \citenamefont {Saw},\ and\ \citenamefont
  {Homann}}]{Bec2016}%
  \BibitemOpen
  \bibfield  {author} {\bibinfo {author} {\bibfnamefont {J.}~\bibnamefont
  {Bec}}, \bibinfo {author} {\bibfnamefont {S.~S.}\ \bibnamefont {Ray}},
  \bibinfo {author} {\bibfnamefont {E.~W.}\ \bibnamefont {Saw}}, \ and\
  \bibinfo {author} {\bibfnamefont {H.}~\bibnamefont {Homann}},\ }\bibfield
  {title} {\enquote {\bibinfo {title} {Abrupt growth of large aggregates by
  correlated coalescences in turbulent flow},}\ }\href@noop {} {\bibfield
  {journal} {\bibinfo  {journal} {Phys. Rev. E}\ }\textbf {\bibinfo {volume}
  {93}},\ \bibinfo {pages} {031102(R)} (\bibinfo {year} {2016})}\BibitemShut
  {NoStop}%
\bibitem [{\citenamefont {Bird}\ \emph {et~al.}(1977)\citenamefont {Bird},
  \citenamefont {Curtiss}, \citenamefont {Armstrong},\ and\ \citenamefont
  {Hassager}}]{Bird}%
  \BibitemOpen
  \bibfield  {author} {\bibinfo {author} {\bibfnamefont {R.~B.}\ \bibnamefont
  {Bird}}, \bibinfo {author} {\bibfnamefont {C.~F.}\ \bibnamefont {Curtiss}},
  \bibinfo {author} {\bibfnamefont {R.~C.}\ \bibnamefont {Armstrong}}, \ and\
  \bibinfo {author} {\bibfnamefont {O.}~\bibnamefont {Hassager}},\ }\href@noop
  {} {\emph {\bibinfo {title} {{Dynamics of Polymeric Liquids}}}}\ (\bibinfo
  {publisher} {John Wiley and Sons},\ \bibinfo {address} {New York},\ \bibinfo
  {year} {1977})\BibitemShut {NoStop}%
\bibitem [{\citenamefont {Perlekar}\ and\ \citenamefont
  {Pandit}(2011)}]{prasad2011}%
  \BibitemOpen
  \bibfield  {author} {\bibinfo {author} {\bibfnamefont {P.}~\bibnamefont
  {Perlekar}}\ and\ \bibinfo {author} {\bibfnamefont {R.}~\bibnamefont
  {Pandit}},\ }\bibfield  {title} {\enquote {\bibinfo {title} {Statistically
  steady turbulence in thin films: direct numerical simulations with {E}kman
  friction},}\ }\href@noop {} {\bibfield  {journal} {\bibinfo  {journal} {New
  J. Phys.}\ }\textbf {\bibinfo {volume} {11}},\ \bibinfo {pages} {073003}
  (\bibinfo {year} {2011})}\BibitemShut {NoStop}%
\bibitem [{\citenamefont {Boffetta}\ and\ \citenamefont
  {Ecke}(2012)}]{Boffetta-Ann-Rev}%
  \BibitemOpen
  \bibfield  {author} {\bibinfo {author} {\bibfnamefont {G.}~\bibnamefont
  {Boffetta}}\ and\ \bibinfo {author} {\bibfnamefont {R.~E.}\ \bibnamefont
  {Ecke}},\ }\bibfield  {title} {\enquote {\bibinfo {title} {Two-dimensional
  turbulence},}\ }\href {\doibase 10.1146/annurev-fluid-120710-101240}
  {\bibfield  {journal} {\bibinfo  {journal} {Annu. Rev. Fluid Mech.}\ }\textbf
  {\bibinfo {volume} {44}},\ \bibinfo {pages} {427--451} (\bibinfo {year}
  {2012})}\BibitemShut {NoStop}%
\bibitem [{\citenamefont {Jin}\ and\ \citenamefont
  {Collins}(2007)}]{Collins2007}%
  \BibitemOpen
  \bibfield  {author} {\bibinfo {author} {\bibfnamefont {S.}~\bibnamefont
  {Jin}}\ and\ \bibinfo {author} {\bibfnamefont {L.~R.}\ \bibnamefont
  {Collins}},\ }\bibfield  {title} {\enquote {\bibinfo {title} {Dynamics of
  dissolved polymer chains in isotropic turbulence},}\ }\href@noop {}
  {\bibfield  {journal} {\bibinfo  {journal} {New J. Phys.}\ }\textbf {\bibinfo
  {volume} {9}},\ \bibinfo {pages} {360} (\bibinfo {year} {2007})}\BibitemShut
  {NoStop}%
\bibitem [{Youtube movie showing the evolution of inertia-less elastic chains:
  \url{https://youtu.be/etLuK6ovAqk}()}]{Youtube_tr}%
  \BibitemOpen
  Youtube movie showing the evolution of inertia-less elastic chains:
  \url{https://youtu.be/etLuK6ovAqk},\ \href@noop {} {}\BibitemShut {NoStop}%
\bibitem [{Youtube movie showing the evolution of heavy-headed chains:
  \url{https://www.youtube.com/watch?v=DG3TJTye2_8}()}]{Youtube_hh}%
  \BibitemOpen
  Youtube movie showing the evolution of heavy-headed chains:
  \url{https://www.youtube.com/watch?v=DG3TJTye2_8},\ \href@noop {}
  {}\BibitemShut {NoStop}%
\bibitem [{Youtube movie showing the evolution of uniformly-inertial chains:
  \url{https://youtube.com/watch?v=g0T6JvVVI20}()}]{Youtube_ui}%
  \BibitemOpen
  Youtube movie showing the evolution of uniformly-inertial chains:
  \url{https://youtube.com/watch?v=g0T6JvVVI20},\ \href@noop {} {}\BibitemShut
  {NoStop}%
\bibitem [{\citenamefont {Mitra}\ and\ \citenamefont
  {Perlekar}(2018)}]{Dhruba2018}%
  \BibitemOpen
  \bibfield  {author} {\bibinfo {author} {\bibfnamefont {D.}~\bibnamefont
  {Mitra}}\ and\ \bibinfo {author} {\bibfnamefont {P.}~\bibnamefont
  {Perlekar}},\ }\bibfield  {title} {\enquote {\bibinfo {title} {Topology of
  two-dimensional turbulent flows of dust and gas},}\ }\href {\doibase
  10.1103/PhysRevFluids.3.044303} {\bibfield  {journal} {\bibinfo  {journal}
  {Phys. Rev. Fluids}\ }\textbf {\bibinfo {volume} {3}},\ \bibinfo {pages}
  {044303} (\bibinfo {year} {2018})}\BibitemShut {NoStop}%
\bibitem [{\citenamefont {Gupta}\ \emph {et~al.}(2015)\citenamefont {Gupta},
  \citenamefont {Perlekar},\ and\ \citenamefont {Pandit}}]{Gupta2015}%
  \BibitemOpen
  \bibfield  {author} {\bibinfo {author} {\bibfnamefont {A.}~\bibnamefont
  {Gupta}}, \bibinfo {author} {\bibfnamefont {P.}~\bibnamefont {Perlekar}}, \
  and\ \bibinfo {author} {\bibfnamefont {R.}~\bibnamefont {Pandit}},\
  }\bibfield  {title} {\enquote {\bibinfo {title} {Two-dimensional homogeneous
  isotropic fluid turbulence with polymer additives},}\ }\href {\doibase
  10.1103/PhysRevE.91.033013} {\bibfield  {journal} {\bibinfo  {journal} {Phys.
  Rev. E}\ }\textbf {\bibinfo {volume} {91}},\ \bibinfo {pages} {033013}
  (\bibinfo {year} {2015})}\BibitemShut {NoStop}%
\bibitem [{\citenamefont {Breier}\ \emph {et~al.}(2018)\citenamefont {Breier},
  \citenamefont {Lalescu}, \citenamefont {Waas}, \citenamefont {Wilczek},\ and\
  \citenamefont {Mazza}}]{Breier2018}%
  \BibitemOpen
  \bibfield  {author} {\bibinfo {author} {\bibfnamefont {R.~E.}\ \bibnamefont
  {Breier}}, \bibinfo {author} {\bibfnamefont {C.~C.}\ \bibnamefont {Lalescu}},
  \bibinfo {author} {\bibfnamefont {D.}~\bibnamefont {Waas}}, \bibinfo {author}
  {\bibfnamefont {M.}~\bibnamefont {Wilczek}}, \ and\ \bibinfo {author}
  {\bibfnamefont {M.~G.}\ \bibnamefont {Mazza}},\ }\bibfield  {title} {\enquote
  {\bibinfo {title} {Emergence of phytoplankton patchiness at small scales in
  mild turbulence},}\ }\href {\doibase 10.1073/pnas.1808711115} {\bibfield
  {journal} {\bibinfo  {journal} {Proc. Nat. Acad. Sci. USA}\ }\textbf
  {\bibinfo {volume} {115}},\ \bibinfo {pages} {12112--12117} (\bibinfo {year}
  {2018})}\BibitemShut {NoStop}%
\bibitem [{\citenamefont {Gupta}\ \emph {et~al.}(2018)\citenamefont {Gupta},
  \citenamefont {Roy}, \citenamefont {Saha},\ and\ \citenamefont
  {Ray}}]{SSRay2018}%
  \BibitemOpen
  \bibfield  {author} {\bibinfo {author} {\bibfnamefont {A.}~\bibnamefont
  {Gupta}}, \bibinfo {author} {\bibfnamefont {A.}~\bibnamefont {Roy}}, \bibinfo
  {author} {\bibfnamefont {A.}~\bibnamefont {Saha}}, \ and\ \bibinfo {author}
  {\bibfnamefont {S.~S.}\ \bibnamefont {Ray}},\ }\bibfield  {title} {\enquote
  {\bibinfo {title} {Flocking of active particles in a turbulent flow},}\
  }\href@noop {} {\bibfield  {journal} {\bibinfo  {journal} {ArXiv e-prints}\ }
  (\bibinfo {year} {2018})},\ \Eprint {http://arxiv.org/abs/1812.10288}
  {arXiv:1812.10288} \BibitemShut {NoStop}%
\bibitem [{\citenamefont {Choudhary}\ \emph {et~al.}(2015)\citenamefont
  {Choudhary}, \citenamefont {Venkataraman},\ and\ \citenamefont
  {Ray}}]{Choudhary2015}%
  \BibitemOpen
  \bibfield  {author} {\bibinfo {author} {\bibfnamefont {A.}~\bibnamefont
  {Choudhary}}, \bibinfo {author} {\bibfnamefont {D.}~\bibnamefont
  {Venkataraman}}, \ and\ \bibinfo {author} {\bibfnamefont {S.~S.}\
  \bibnamefont {Ray}},\ }\bibfield  {title} {\enquote {\bibinfo {title} {Effect
  of inertia on model flocks in a turbulent environment},}\ }\href {\doibase
  10.1209/0295-5075/112/24005} {\bibfield  {journal} {\bibinfo  {journal}
  {Europhys. Lett.}\ }\textbf {\bibinfo {volume} {112}},\ \bibinfo {pages}
  {24005} (\bibinfo {year} {2015})}\BibitemShut {NoStop}%
\end{thebibliography}%
\end{document}